\documentclass[%
 reprint,
nofootinbib,
 amsmath,amssymb,
 aps,
 pra,
]{revtex4-2}

\usepackage{graphicx,subfigure}
\usepackage{dcolumn}
\usepackage{bm}
\usepackage[T1]{fontenc}
\usepackage[section]{placeins}
\usepackage{float}
\usepackage{hyperref}

\usepackage{color}
\definecolor{red}{rgb}{0.9,0,0}
\definecolor{magenta}{rgb}{1.0,0,1.0}

\begin{document}

\title{An application benchmark for fermionic quantum simulations}

\author{Pierre-Luc Dallaire-Demers}
\email{pierre-luc@zapatacomputing.com}
\affiliation{Zapata Computing Inc.}
\author{Micha\l{} St\k{e}ch\l{}y}
\affiliation{Zapata Computing Inc.}
\author{Jerome F. Gonthier}
\affiliation{Zapata Computing Inc.}
\author{Ntwali Toussaint Bashige}
\affiliation{Zapata Computing Inc.}
\author{Jonathan Romero}
\affiliation{Zapata Computing Inc.}
\author{Yudong Cao}
\email{yudong@zapatacomputing.com}
\affiliation{Zapata Computing Inc.}

\date{\today}

\begin{abstract}
It is expected that the simulation of correlated fermions in chemistry and material science will be one of the first practical applications of quantum processors. 
Given the rapid evolution of quantum hardware, it is increasingly important to develop robust benchmarking techniques to gauge the capacity of quantum hardware specifically for the purpose of fermionic simulation.
Here we propose using the one-dimensional Fermi-Hubbard model as an application benchmark for variational quantum simulations on near-term quantum devices. Since the one-dimensional Hubbard model is both strongly correlated and exactly solvable with the Bethe ansatz, it provides a reference ground state energy that a given device with limited coherence will be able to approximate up to a maximal size. The length of the largest chain that can be simulated provides an \emph{effective fermionic length}. We use variational quantum eigensolver to approximate the ground state energy values of Fermi-Hubbard instances and
show how the fermionic length benchmark can be used in practice to assess the performance of bounded-depth devices in a scalable fashion.
\end{abstract}

\maketitle

\section{Introduction \label{sec:Introduction}}

The study of quantum algorithms for Noisy Intermediate-Scale Quantum (NISQ) \cite{Preskill2018} computers is an important development in recent years triggered by the rapid evolution of quantum hardware. These NISQ algorithms are intended for specific applications such as simulating quantum systems \cite{Peruzzo2014,Wecker2015,McClean2016,Kandala2017,Romero2018,Dallaire2019,McArdle2019,Yuan2019}, combinatorial optimization \cite{FarhiQAOA,zhou2018quantum}, machine learning \cite{schuld2018circuitcentric,Dallaire2018,Havlek2019,Schuld2019,Benedetti2019} and more \cite{xu2019variational,bravoprieto2019variational,huang2019nearterm}. Quantum circuits arising from these algorithms are often structured to account for the specific characters of the application problem. For instance, the variational quantum eigensolver (VQE) \cite{Peruzzo2014,Wecker2015,McClean2016,Kandala2017,Romero2018,Dallaire2019} has specific structures built into circuit ansatzes to account for the unique properties of interacting fermions \cite{Romero2018,Dallaire2019}.

With an increasingly diverse set of hardware devices available, it remains unclear how best to benchmark the usefulness of devices for the specific purpose of variational fermionic simulation. Current benchmarking techniques, such as randomized benchmarking \cite{Proctor2017}, volumetric benchmark \cite{Cross2019} or cross-entropy benchmark \cite{Arute2019}, do not take into account
the structured nature of quantum circuits arising from specific NISQ algorithms. As a result, for a given NISQ algorithm, it is unclear how the circuit structure and the hardware limitations affect the performance of a device for the specific application. To address this issue in the context of VQE, we consider the ground state problem of the one-dimensional Fermi-Hubbard model (FHM) as a benchmark. It is an exactly solvable model by the Bethe ansatz for both finite and infinite chains. In addition, it is a relatively simple model that nonetheless captures the essential complexity of preparing the ground state and simulating strongly correlated fermionic systems.

There are certain generic features that are expected from such a benchmark (Fig.\ \ref{fig:TheoreticalPerformanceCurve}). For preparing the ground state of an $L$-site FHM on a quantum computer, both the number of qubits and circuit depth are expected to grow as $L$ increases. Assuming adequate training of the circuit ansatz, the final ground state energy obtained from the VQE calculations should decrease with $L$, approaching the infinite-chain limit $E_0$. For a given NISQ device that is limited to running shallow circuits, however, after a certain chain length $L^*$ the ground state energy obtained is expected to diverge from the infinite-chain limit. In other words, there is a turning point that serves as an \emph{effective fermionic length} that characterizes the capacity of the quantum device for fermionic simulation. This way one can precisely delineate the interplay between the circuit structure (which may manifest as circuit depth and number of gates) and the physical hardware limitations (which may manifest as gate error and decoherence). 

As an example, we consider the details of such benchmark for the quantum processor recently produced by the team at Google (the \emph{Sycamore} processor \cite{Arute2019}). 
By parametrizing the pulses used to operate the tunable couplers and the qubit frequencies, it is possible to use the Sycamore device as a variational ansatz. It has been demonstrated experimentally that variational 2-qubit gates can be implemented \cite{Foxen2020}. Each parametrized two-qubit gate has two components---an exchange term and a tunable dispersive interaction. We define a practical variational building block by starting each step with a variable X rotation to select a basis and by adding tunable Z phases at the end of each step to compensate for stray phase-shifts. A variational layer is composed of many parallel 2-qubit elements which are parametrized such that the experimental implementation of each layer is completed in a fixed time. This allows a simple multi-layer composition of the ansatz.
By construction the ansatz can interpolate between discrete elements of the class of random circuits used for the supremacy demonstration \cite{Arute2019}.
All single-qubit gates can be reached by the ansatz as well as two-qubit $\textrm{cphase}$ operations and non-nearest-neighbor matchgates \cite{Brod2013} which are both universal for quantum computing. In general, we expect the ansatz to be hard to simulate classically (except for certain special cases\footnote{For example, consider a circuit of nearest-neighbor iSWAP gates acting only on a chain of qubits. Since iSWAP is a matchgate, such circuit can be classically simulated efficiently \cite{Terhal2002,Jozsa2008}}) and can also be used to represent and study fermionic states beyond the reach of classical computers.

The paper is organized as follows: In Section \ref{sec:Benchmark} we describe how the one-dimensional Fermi-Hubbard model can be used as a benchmark for fermionic ansatz on devices that operate beyond the supremacy regime. In Section \ref{sec:Implementation} we describe a heuristic to optimize a VQE ansatz layer-by-layer \cite{Benedetti2019} in order to mitigate the barren plateau problem, along with a proposal for a hardware-efficient ansatz for the Sycamore device and specify general architecture constraints for other hardware architectures. As a proof-of-concept we numerically simulate how the benchmark would perform on the Sycamore device in Section \ref{sec:Numerics}. 
Finally, in Section \ref{sec:Discussion} we discuss results and possible future lines of work.

\section{The benchmark \label{sec:Benchmark}}

\begin{figure}[tp]
\includegraphics[width=8cm]{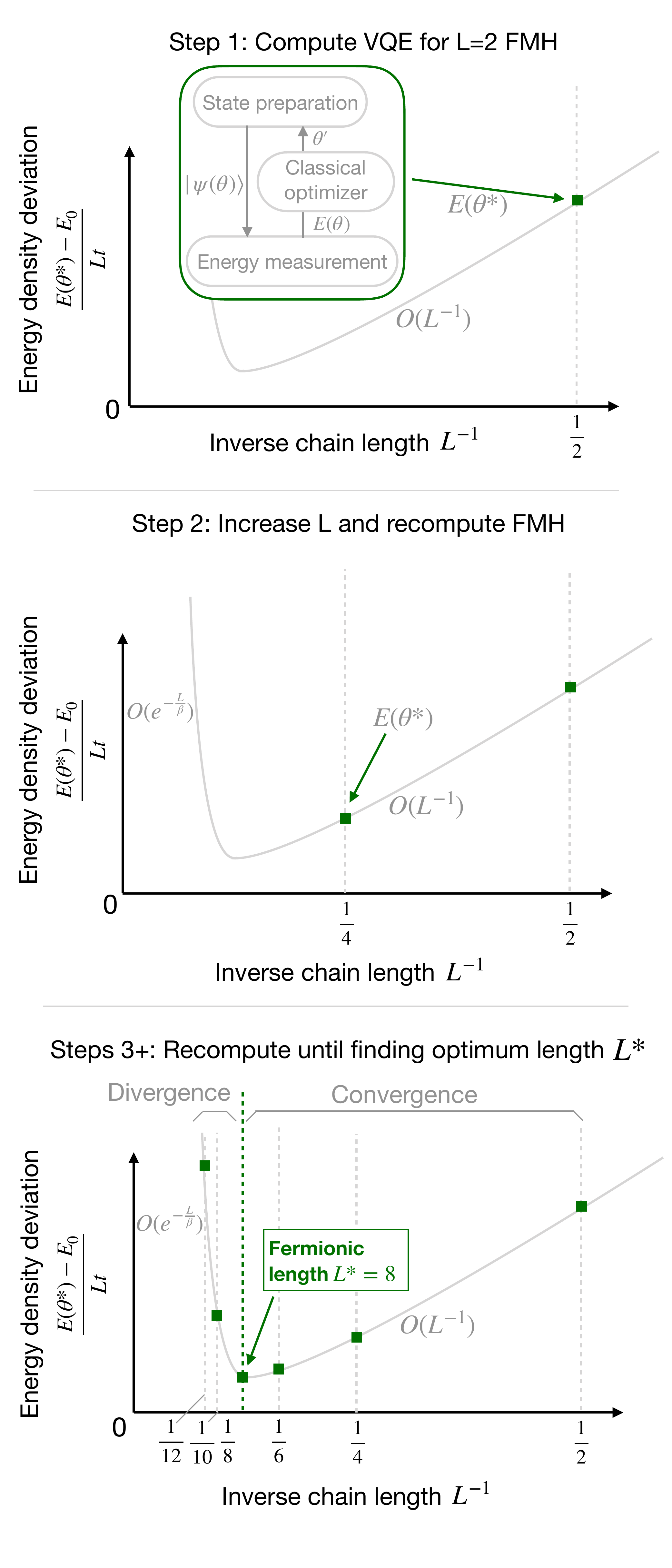}
\caption{
Schematics for the application benchmark scheme (data points are for illustration purposes only). The goal is to generate the energy density deviation $\frac{E(\theta^*)-E_0}{Lt}$ as a function of inverse chain length $L^{-1}$. Here $E(\theta^*)$ values come from VQE calculations, $E_0$ is the infinite-chain ground state energy for 1D FHM (see Eq.~\eqref{eq:exactenergydensity}) and $t$ is the hopping potential (see Eq.~\eqref{eq:onedimensionalFHM}). 
The benchmarking process starts with a short chain and proceeds to increment $L$ while extracting $E(\theta^*)$ on the quantum device. As $L$ increases, $E(\theta^*)$ should converge towards the infinite chain limit $E_0$. However, due to noise and decoherence, after some value $L^*$, the estimated ground state energy density will diverge away from the infinite-chain limit, yielding a scaling of $O(e^{-\frac{L}{\beta}})$ in the energy density deviation. Here $\beta$ is a parameter depending on the noise and decoherence of the device.
}
\label{fig:TheoreticalPerformanceCurve}
\end{figure}

In this section, we revisit some of the details of the one-dimensional Fermi-Hubbard model (FHM) and describe how it can be used as a benchmark to characterize the performance of a quantum device for the task of simulating fermionic systems with the variational quantum eigensolver algorithm \cite{Peruzzo2014,McClean2016}. Specifically, we show that an \emph{effective fermionic length} (EFL) can be obtained from estimates of the energy density for 1D FHM of increasing size. This metric serves as an estimate of the effective size of fermionic systems that can be simulated using a quantum device and a particular choice of variational circuit ansatz. Fig. \ref{fig:TheoreticalPerformanceCurve} depicts the steps to carry out the estimation of the EFL on an actual device. The rest of the section describes in detail the different aspects of the benchmark.

\subsection{Representation of the one-dimensional FHM \label{sub:Representation1DFHM}}

The one-dimensional FHM describes a physical system of fermions dwelling on a linear chain of $L$ sites. Each site has the capacity for holding at most one fermion which can be either spin up $\uparrow$ or down $\downarrow$. Hence there are in total $2L$ spin orbitals for the entire system, with each site associated with two spin orbitals.
The second quantization formulation of the 1D FHM is given by the Hamiltonian \cite{Bedurftig1997}
\begin{equation}
    \begin{array}{rcl}
    H\left(t,U\right) &=& \displaystyle -t\sum_{\sigma=\uparrow,\downarrow}\sum_{j=1}^{L-1}\left(a_{j+1,\sigma}^{\dagger}a_{j,\sigma}+a_{j,\sigma}^{\dagger}a_{j+1,\sigma}\right)\\
    &&\displaystyle
    +U\sum_{j=1}^{L}n_{j,\uparrow}n_{j,\downarrow}-\mu\sum_{\sigma=\uparrow,\downarrow}\sum_{j=1}^{L}n_{j,\sigma},
    \end{array}
    \label{eq:onedimensionalFHM}
\end{equation}
where the $a_{j,\sigma}$ and $a_{j,\sigma}^{\dagger}$ are creation and annihilation operators for a fermion with spin $\sigma$ at site $j$ and $n_{j,\sigma}=a_{j,\sigma}^{\dagger}a_{j,\sigma}$ are the number operators. This describes electrons hopping along a flat band with $L$ sites with kinetic energy $t$. There is a local Coulomb interaction $U$ and a chemical potential $\mu$ that determines the total number of electrons $N$.  At half-filling we have the total number of electrons $N=L$. The exact ground state energy per site of an infinite chain of the 1D FHM as computed from the Bethe ansatz \cite{Lieb2003, Balzer2008} is given by
\begin{equation}
    \frac{E_0}{Lt}=-4\int_0^{\infty}d\omega \frac{J_0\left(\omega\right)J_1\left(\omega\right)}{\omega \left(1+e^{\frac{\omega U}{2t}}\right)}
    \label{eq:exactenergydensity}
\end{equation}
where $J_0\left(x\right)$ and $J_1\left(x\right)$ are Bessel functions. After numerical integration we tabulate the energy density for a few values of the interaction energy $U$ (see Table \ref{tab:AsymptotiEnergyDensity}).
\begin{table}
\begin{center}
    \begin{tabular}{|c|c|c|c|c|}
    \hline
    &&&&\\[-1em]
    $\frac{U}{t}$&0&2&4&8\\[-1em]
    &&&&\\
    \hline
    &&&&\\[-1em]
    $\frac{E_0}{L}$&-1.27324 &-0.844374 &-0.573729 &-0.327531 \\[-1em]
    &&&&\\
    \hline
    \end{tabular}
\end{center}
\caption{Density of energy $\frac{E_0}{L}$ in the thermodynamic limit of a one-dimensional FHM as a function of the dimensionless interaction energy $\frac{U}{t}$.}
\label{tab:AsymptotiEnergyDensity}
\end{table}
\begin{figure}[tp]
\centering
\includegraphics[width=7cm]{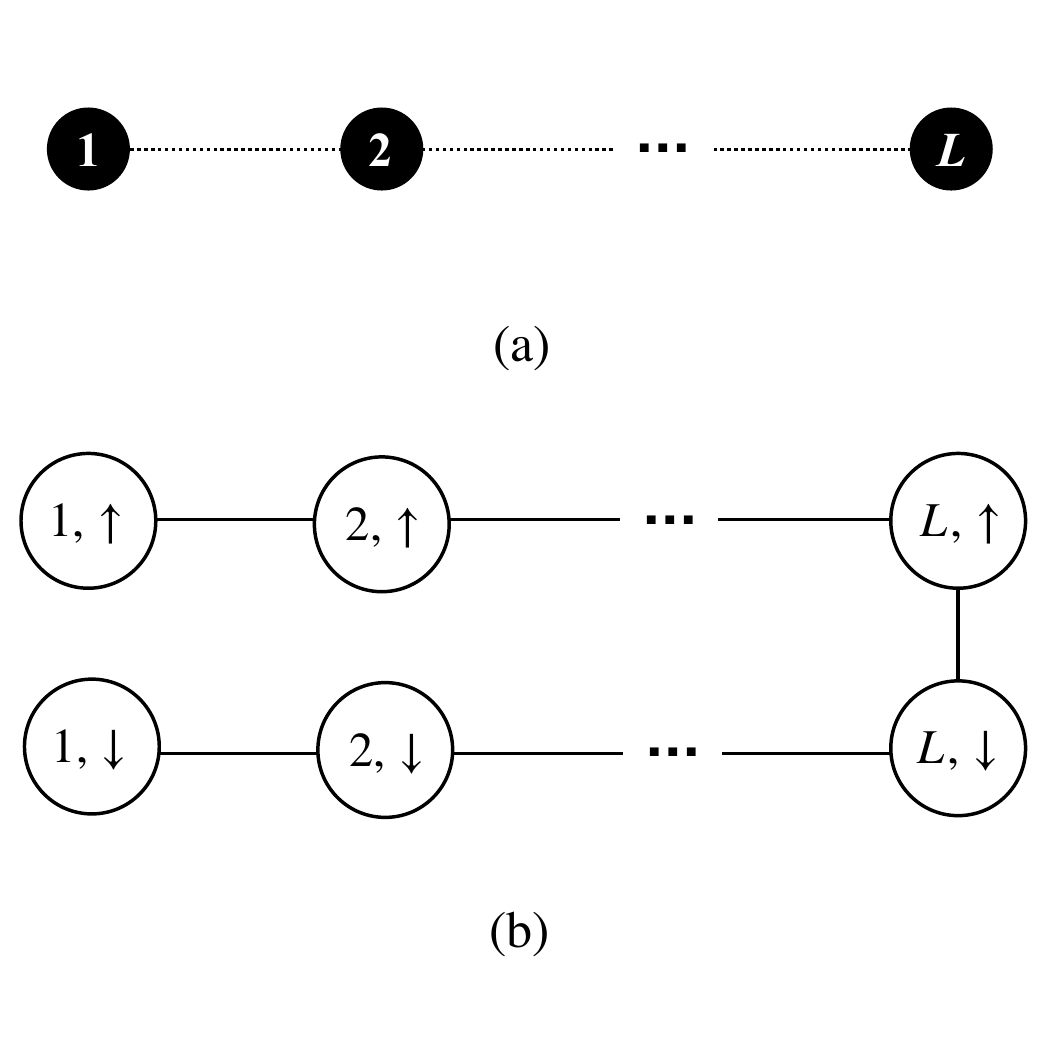}
\caption{(a) One-dimensional chain of sites $1,2,\cdots,L$ in the Fermi-Hubbard model. (b) Canonical Jordan-Wigner encoding for fermions in $2L$ spin-$\frac{1}{2}$ orbitals. Each spin orbital is mapped to a qubit and the edges indicate the qubit connectivity.}
\label{fig:JordanWignerEncoding}
\end{figure}
For a finite chain of length $L$, the correction to the energy density $\frac{E_0}{L}$ is of order $O\left(\frac{1}{L}\right)$ \cite{Essler2005}. Hence the energy density of the ground state of increasingly longer chains will asymptotically reach the limit shown in Table \ref{tab:AsymptotiEnergyDensity}.
In order to map the Hamiltonian of the FHM to a quantum computer, it is useful to introduce the intermediate notation of Majorana fermions $\gamma_{j,\sigma}^A=a_{j,\sigma}^{\dagger}+a_{j,\sigma}$ and $\gamma_{j,\sigma}^B=-i\left(a_{j,\sigma}^{\dagger}-a_{j,\sigma}\right)$ such that $\left\{\gamma_{j,\sigma}^{\alpha},\gamma_{k,\sigma'}^{\beta}\right\}=\delta_{jk}\delta_{\sigma\sigma'}\delta_{\alpha\beta}$ and $\left(\gamma_{j,\sigma}^{\alpha}\right)^2=1$. In the Majorana representation, the Hamiltonian of the 1D FHM has the form
\begin{equation}
    \begin{array}{rcl}
    H\left(t,U\right) &=& \displaystyle \frac{it}{2}\sum_{\sigma=\uparrow,\downarrow}\sum_{j=1}^{L-1}\left(\gamma_{j+1,\sigma}^A\gamma_{j,\sigma}^B+\gamma_{j,\sigma}^A\gamma_{j+1,\sigma}^B\right)\\
    &&\displaystyle -\frac{i}{2}\left(\frac{U}{2}-\mu\right)\sum_{\sigma=\uparrow,\downarrow}\sum_{j=1}^{L}\gamma_{j,\sigma}^A\gamma_{j,\sigma}^B\\
    &&\displaystyle -\frac{U}{4}\sum_{j=1}^{L}\gamma_{j,\uparrow}^A\gamma_{j,\uparrow}^B\gamma_{j,\downarrow}^A\gamma_{j,\downarrow}^B+L\left(\frac{U}{4}-\mu\right),
    \end{array}
    \label{eq:onedimensionalFHM_majorana}
\end{equation}
where half-filling is obtained by setting $\mu = \frac{U}{2}$.

To complete the mapping of the spin orbitals to a set of distinguishable qubits we define and use a canonical Jordan-Wigner encoding \cite{Peruzzo2014} shown in Fig.\ \ref{fig:JordanWignerEncoding}. From the Majorana representation this corresponds to the mapping $Z_{j,\sigma}=-i\gamma_{j,\sigma}^A \gamma_{j,\sigma}^B$ and $X_{j,\sigma}X_{j+1,\sigma}+Y_{j,\sigma}Y_{j+1,\sigma}=i\left(\gamma_{j+1,\sigma}^A\gamma_{j,\sigma}^B+\gamma_{j,\sigma}^A\gamma_{j+1,\sigma}^B\right)$. Here $X$, $Y$, $Z$ are Pauli operators. In the Jordan-Wigner representation the Hamiltonian is written as
\begin{equation}
    \begin{array}{rcl}
    H\left(t,U\right) &=& \displaystyle \frac{t}{2}\sum_{\sigma=\uparrow,\downarrow}\sum_{j=1}^{L-1}\left(X_{j,\sigma}X_{j+1,\sigma}+Y_{j,\sigma}Y_{j+1,\sigma}\right)\\
    &&\displaystyle+\frac{1}{2}\left(\frac{U}{2}-\mu\right)\sum_{\sigma=\uparrow,\downarrow}\sum_{j=1}^{L}Z_{j,\sigma}\\
    &&\displaystyle+\frac{U}{4}\sum_{j=1}^{L}Z_{j,\uparrow}Z_{j,\downarrow}+L\left(\frac{U}{4}-\mu\right).
    \end{array}
    \label{eq:onedimensionalFHM_jordanwigner}
\end{equation}

This particular Jordan-Wigner mapping has the advantage of casting all terms in the Hamiltonian of the 1D FHM (Eq.\ \ref{eq:onedimensionalFHM_majorana}) to operators acting on at most two qubits on quantum processor.

\subsection{The variational energy minimization task \label{sub:VQETask}}

Here we briefly formulate the general steps in the task of finding the ground state of the 1D FHM on a quantum processor. We assume that a parametrized ansatz state $\left|\psi(\vec{\theta})\right\rangle=U(\vec{\theta})\left|0\right\rangle^{\otimes n}$ (over $n\geqslant 2L$ qubits) can be constructed for an arbitrary assignment of parameters $\vec{\theta}$. We will denote the expectation value of an operator $O$ as $\left\langle O\right\rangle_{\vec{\theta}}=\textrm{tr}\left( O \left|\psi(\vec{\theta})\right\rangle \left\langle\psi(\vec{\theta})\right| \right)$. The variational energy of the 1D FHM at half-filling is simply the statistical expectation value
\begin{equation}
    \begin{array}{rcl}
         E(\vec{\theta})&=& \displaystyle \left\langle H(t,U)\right\rangle_{\vec{\theta}} + \mu \left\langle N\right\rangle_{\vec{\theta}}\\
         &=& \displaystyle \frac{t}{2}\sum_{\sigma=\uparrow,\downarrow}\sum_{j=1}^{L-1}\left(\left\langle X_{j,\sigma}X_{j+1,\sigma}\right\rangle_{\vec{\theta}}+\left\langle Y_{j,\sigma}Y_{j+1,\sigma}\right\rangle_{\vec{\theta}}\right)\\
         &&\displaystyle+\frac{U}{4}\sum_{j=1}^L\left(\left\langle Z_{j,\uparrow}Z_{j,\downarrow}\right\rangle_{\vec{\theta}}+1\right).
    \end{array}
    \label{eq:onedimensionalFHM_statistics}
\end{equation}
The task of finding the ground state corresponds to the optimization problem $E(\vec{\theta}^*)=\textrm{min}_{\vec{\theta}} E(\vec{\theta})$. The optimal energy, $E(\vec{\theta}^*)$ computed for a specific chain length $L$, provides an estimate of the infinite chain energy density with deviation $\frac{E(\vec{\theta}^*)-E_0}{Lt}$, where $E_0$ is obtained analytically from Eq.~\eqref{eq:exactenergydensity}.
It can be seen that $E(\vec{\theta})$ can be evaluated by sampling in only three different measurement bases. 
Let us remark that the number of measurements scales as $O(\frac{1}{L\epsilon^2})$ for error $\epsilon$ in estimating the chain energy density \cite{Wecker2015}. This implies that for a fixed $\epsilon$ it requires \emph{fewer} measurements as the chain gets longer.

\subsection{Definition of the benchmark \label{sub:BenchmarkDefinition}}

A quantum computer with the ability to simulate accurately the 1D FHM would be able to provide estimates of the energy per site in the thermodynamic limit with accuracy $O(\frac{1}{L})$, as described in the previous section. Such a simulation can be carried out using the VQE algorithm implemented with a hardware-efficient ansatz. This ansatz type generally consists of a series of repeating parameterized quantum circuits, typically referred to as ``layers'' \cite{Kandala2017,cao2019quantum}. We anticipate that the number of layers in the ansatz required to describe the energy accurately increases polynomially with the size of the chain\footnote{The task of preparing the optimal mean-field approximation to the ground state of the 1D FHM on a quantum computer requires a linear-depth circuit of matchgates \cite{Dallaire2019,kivlichan2018,Jiang2018fermion} to prepare the corresponding Gaussian state. This has been shown to be optimal \cite{Jiang2018fermion} for a linear connectivity. Preparing the exact solution requires the preparation of non-Gaussian fermionic states, which can be achieved with circuits of linear depth \cite{Dallaire2019,Jiang2018fermion} incorporating a non-matchgate interaction. Therefore, we hypothesize that at least a linear-depth circuit would be required to achieve such preparation. The variational non-Gaussian space also provides a setting to prepare fermionic state from other strongly correlated models. Reference \cite{Dallaire2019} provides some numerical evidence supporting this hypothesis.}. In a NISQ device, increasing the number of layers decreases the overall fidelity of the variational circuit, introducing an error $E(\vec{\theta}^*)$ in the energy estimate obtained with VQE that grows with the chain length $L$.
Consequently, the overall error in the infinite size energy density computed by VQE will be the result of the interplay between the errors associated to a finite chain size, which scales as $O(\frac{1}{L})$, and the error introduced by decoherence, which increases as $L\rightarrow\infty$, as illustrated in Fig.\ \ref{fig:TheoreticalPerformanceCurve}. Therefore, the overall error in the estimate of the energy density will follow a behavior as the one described by the gray line in Fig.\ \ref{fig:TheoreticalPerformanceCurve}, where the best approximation to the infinite chain energy density is achieved for
a length $L^*$. We call this quantity $L^*$ the \emph{effective fermionic length} (EFL), corresponding to the maximum length of a 1D FHM for which a quantum device implementing a hardware-efficient ansatz provides the best estimate of the infinite chain energy density. The accuracy of this estimate will also depend on the performance of the VQE procedure, which is influenced by the quality of the VQE optimization. In this sense, $L^*$ can be interpreted as a holistic metric describing both the power of a quantum device as an ansatz for simulating fermionic systems with VQE, and the quality of the VQE procedure itself. We point out in the next section that the VQE optimization for the 1D FHM is likely to be feasible for arbitrary $L$.


In the next section, we also illustrate the implementation of our benchmark with a proposal of a hardware efficient ansatz and VQE optimization strategy to compute the energy of the 1D FHM on a quantum processor. 
We provide implementation details based on the architecture of Google's Sycamore quantum chip, which has been recently used to demonstrate quantum supremacy \cite{Arute2019}.

\section{Benchmark implementation \label{sec:Implementation}}

In this section, we formalize the definition of the class of hardware efficient ansatzes (HEA), and provide a specific example tailored to the simulation of fermionic systems. We provide details on how the Sycamore device can be operated to implement this ansatz.

\subsection{VQE ansatz and optimization strategy \label{sec:GHEA}}

Hardware efficient ansatzes are motivated by the idea of utilizing the native set of parameterizable operations available on a quantum processor. This allows for the design of shallow variational circuits \cite{Kandala2017, Dallaire2019}. We formalize the concept of a hardware efficient ansatz as a parameterized quantum circuit characterized by the following elements:
\begin{enumerate}
	\item A graph of qubits described by an adjacency matrix with maximum degree $K$. The edges designate the physical coupling between the qubits.
	\item A variational two-qubit gate and variational single-qubit elements such that they can be composed to (approximately) generate any unitary transformation on the state of the qubits.
	\item A structure of $K$ layers which can be described as a sequence of sets of edges.
	\item A depth limit mostly constrained by the noise and decoherence of the system.
\end{enumerate}

This description also includes devices with parameterized multi-qubit gates that can be operated in parallel on a number of qubits proportional to the total size of the quantum processor. For example, trapped ion quantum computers \cite{Pino2020} with a lattice of shuttling ions could implement such a variational scheme, where the layers defining the ansatz can include multiqubit operations.

The main challenge of optimizing hardware efficient ansatzes is the existence of barren plateaus in the cost function landscape \cite{Mcclean2018,Cerezo2020}. This phenomenon corresponds to the observation that gradients of cost functions consisting of global operators vanish at a rate that scales as $O(\exp(-n))$ for ansatzes that approximate 2-designs, where $n$ is the total number of qubits. This implies that the strategy of randomly initializing the parameters of hardware efficient ansatzes, many of which might approximate 2-designs, would not be effective for optimization with gradient based approaches. However, it has been shown \cite{Cerezo2020} that shallow circuits (i.e.\ with a depth that scales as $O(\textrm{polylog}(n))$) attain polynomially vanishing gradients for cost functions comprising local operators. This suggests that training shallow parameterized quantum circuits might be possible even with random initialization of the parameters for local cost functions .

Fortunately, the 1D FHM Hamiltonian qualifies as a local cost function, as it consists only of 1-local and 2-local qubit operators (Eq.~\ref{eq:onedimensionalFHM_statistics}). Correspondingly, we expect gradient-based optimization to be feasible for a hardware efficient ansatz with a few number of layers. To take advantage of this, we propose a layer-by-layer optimization strategy to carry out the optimization \cite{Benedetti2019}. We start by optimizing $O(\log(n))$ layers by randomly initializing the parameters. Convergence is achieved after a certain threshold of change in energy between two iterations is met or when a maximum number of function evaluations is reached. This first optimization step provides an approximation to the wavefunction with non-zero overlap with the exact ground state. After completing the first optimization, we increment the number of layers, initializing the new layers according to some random distribution of parameters and retaining the optimal parameters for the old layers. We choose a small interval of angles such that the identity can be recovered but initial symmetries are broken. New layers and the layers from the previous steps are trained using a numerical optimizer. We repeat this procedure until achieving an energy convergence within a predefined global threshold. By doing the optimization sequentially, we approximately guarantee that the starting point for each iteration maintains significant overlap with the ground state.

Finally, we propose a training method which offloads all fermionic gaussian operations to the classical processor and maximizes the use of the variational non-gaussian resource given by the quantum device. The method is described in Appendix \ref{sec:AdvancedTraining}.

\subsection{Example of Sycamore \label{sub:SycamoreExample}}

In the supremacy experiment, the Google team has demonstrated that a microfabricated device can deterministically be put into a state whose statistics cannot be sampled efficiently with a classical computer according to a cross-entropy benchmark \cite{Arute2019}. The space of random circuits used for the supremacy experiment can be made variational by parametrizing the control pulses used to implement the gates on the device. Since the natural time evolution of the tunable couplers of Sycamore corresponds to gates that can be used to implement a fermionic ansatz like the low-depth circuit ansatz \cite{Dallaire2019}, such processors may be naturally suited for fermionic quantum simulations. We propose using the device as an ansatz by doing variational optimization directly on the coupler's parameters and the single qubit detunings. 

The Sycamore device is composed of an array of 54 transmon qubits. Each single-qubit gates can be executed in 25 ns while entangling two-qubit gates can be done in 12 ns. Individual qubits can be measured in the computational basis. The qubits in Sycamore are coupled by tunable couplers. A tunable coupler is essentially another qubit that can be brought into resonance with the neighboring qubits to enable an exchange interaction. This is done by applying a DC flux pulse of a certain amplitude and duration and by controlling the detuning of the qubits.

One of the main advantages of using superconducting circuits as a platform for quantum information processing is that they do not dissipate DC currents. However their dynamical operation can still dissipate energy and the qubit's lifetime and phase coherence are therefore finite. In current technologies, the cross-talk is a significant source of error in the implementation of quantum gates. By operating a device as a variational ansatz, it should be possible to mitigate the effect of cross-talk as well as some systematic coherent errors like under- and over-rotations induced by calibration errors.

To define the hardware efficient ansatz for the Sycamore device, we first choose an ideal variational two-qubit gate which is close to the operational capability of the physical device. We then assign the controls that modulate the ideal variational angles as experimental variational parameters. Finally we choose a structure of layers of variational two-qubit gates that can be used to compose an ansatz of arbitrary depth. 

\begin{figure}[tp]
    \centering
    \includegraphics[width=7cm]{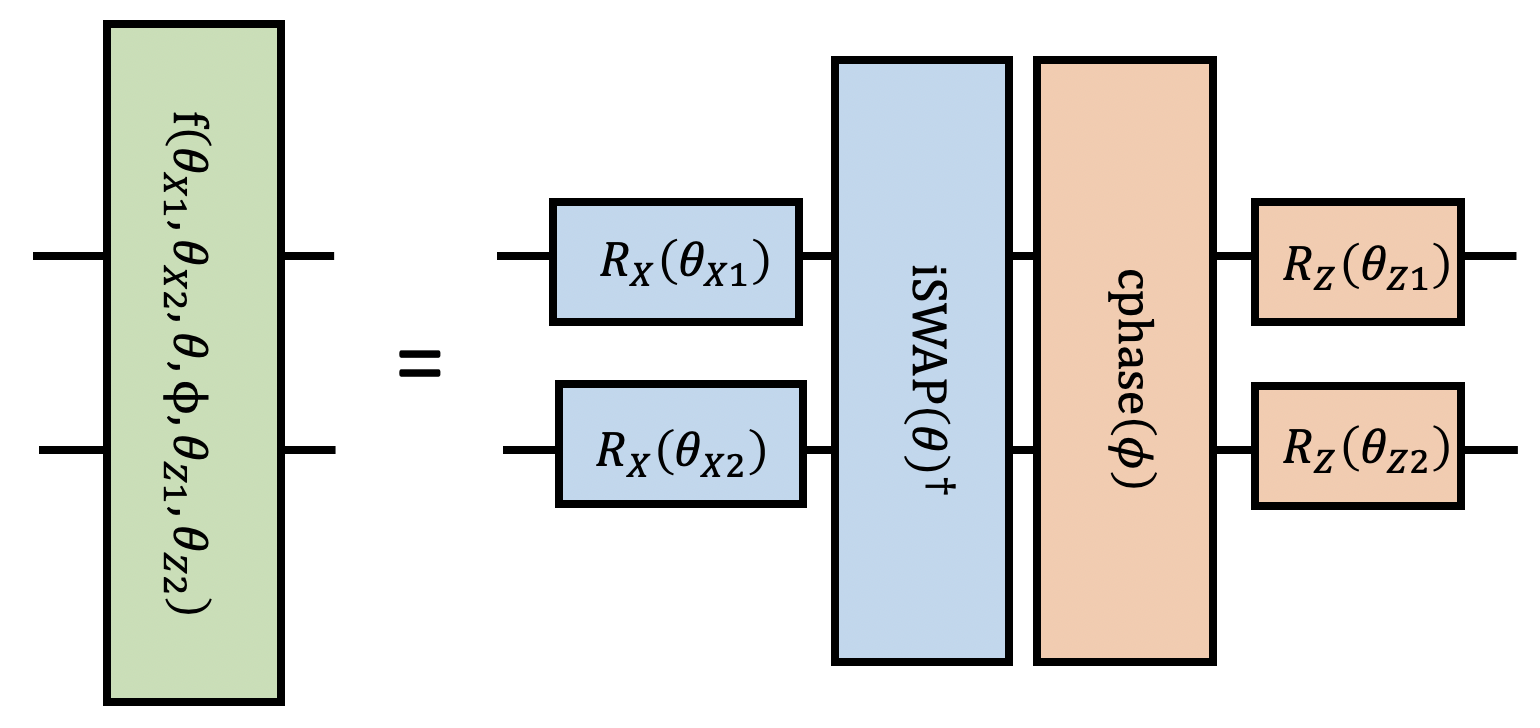}
    \caption{Ideal form of the variational 2-qubit element. The variational gate $f$ starts with two single-qubit $X$ rotations $R_X(\theta_{X1})$ on the first qubit and $R_X(\theta_{X2})$ on the second qubit. It is followed by a variational 2-qubit $\textrm{iSWAP}(\theta)^{\dagger}$ and $\textrm{cphase}(\phi)$. Finally we apply the single-qubit $Z$ rotations $R_Z(\theta_{Z1})$ and $R_Z(\theta_{Z2})$.}
    \label{fig:TwoQubitIdeal}
\end{figure}

From the form of the perturbed iSWAP gate \cite[$f_{\textrm{sim}}$ gate]{Arute2019}, we propose the ideal variational two-qubit gate shown in Fig. \ref{fig:TwoQubitIdeal}. The variational entangling components are chosen to be those naturally implemented by the physics of tunable couplers, namely $\textrm{iSWAP}(\theta)^{\dagger}=e^{-i\theta(X_1 X_2 + Y_1 Y_2)}$ coming from the exchange part of the interaction and $\textrm{cphase}(\phi)=e^{i \phi Z_1 Z_2}$ from the dispersive part. The order of the two-qubit variational gates $XX+YY$ and $ZZ$ can be reversed since they commute. Furthermore, the variational $f$ gate is chosen such that composing two such operations in sequence can be used to generate arbitrary single qubit rotations. This is done by applying single-qubit $X$ rotations $R_X(\theta_{X1})=e^{i \theta_{X1} X_1}$ and $R_X(\theta_{X2})=e^{i \theta_{X2} X_2}$ on the first and second qubit respectively at the beginning of $f$ and by applying the single-qubit $Z$ rotations $R_Z(\theta_{Z1})=e^{i \theta_{Z1} Z_1}$ and $R_Z(\theta_{Z2})=e^{i \theta_{Z2} Z_2}$ at the end of the gate. The $Z$ rotations are also generators for local fermionic Gaussian transformations. The composition of a sequence of variational elements $f$ can be used to construct a set of gates which is universal for quantum computing.

\begin{figure}[tp]
\centering
\subfigure[Parameterization of a tunable coupler.]{\includegraphics[width=8cm]{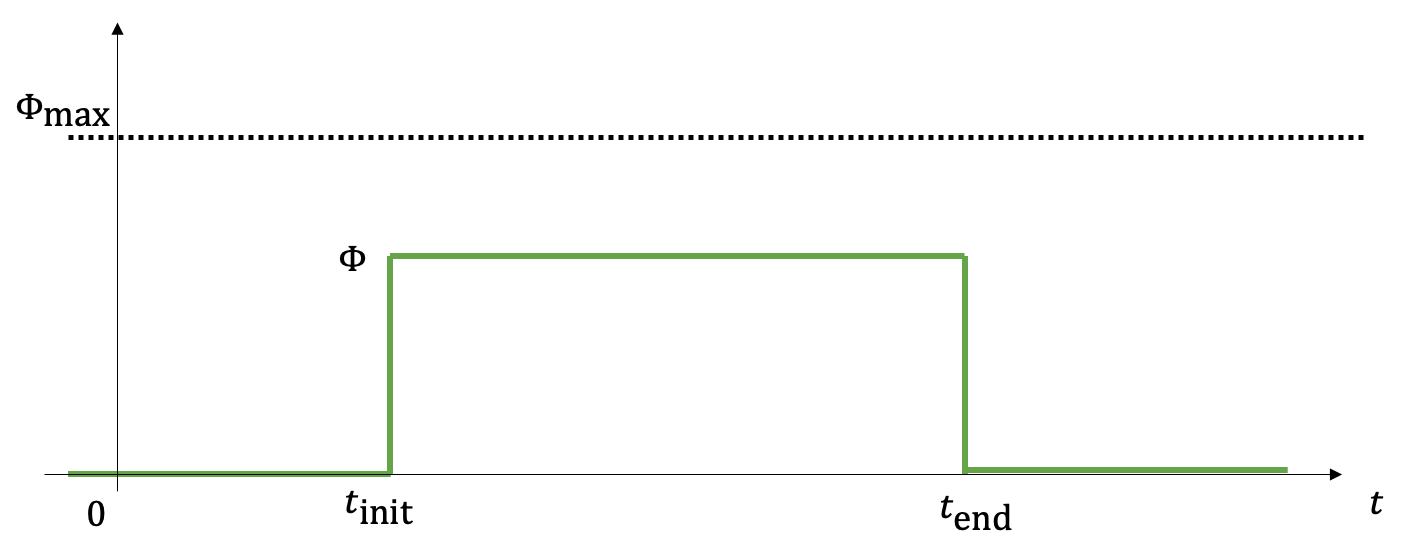}}\\
\subfigure[Physical realization of the tunable two-qubit gate]{\includegraphics[width=8cm]{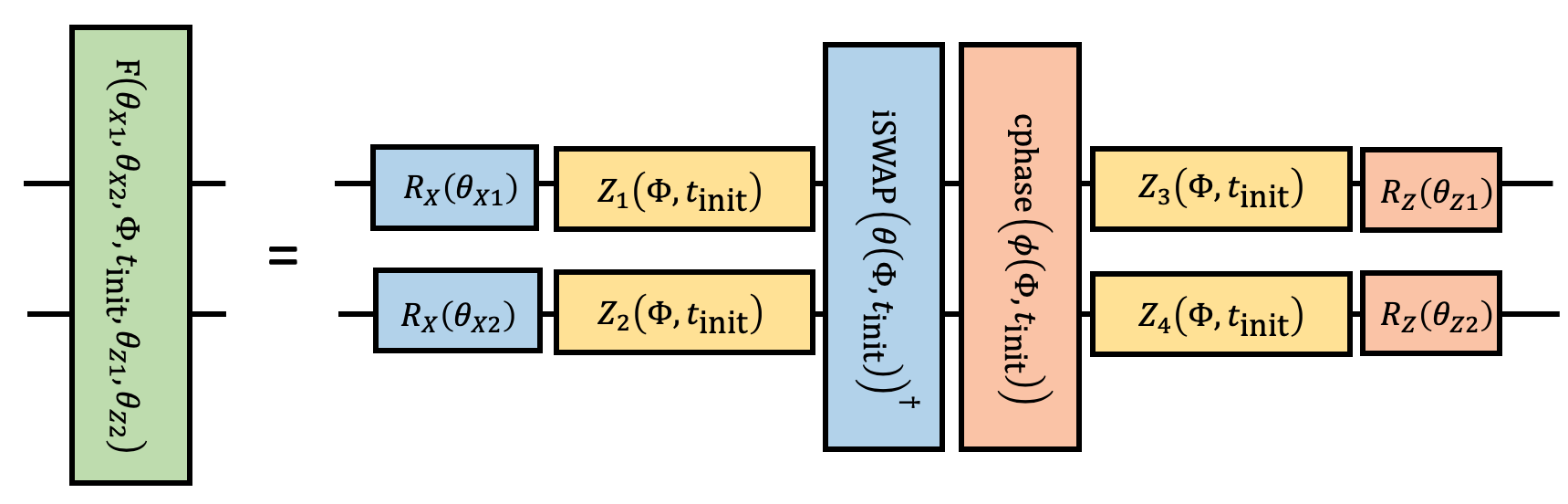}}
\caption{(a) Parametrization of tunable coupler. The amplitude $\Phi$ is parametrized and bounded by $\pm \Phi_{\textrm{max}}$, the maximal value used to define the $f_{\textrm{sim}}$ gate \cite{Arute2019}. The maximal time $t_{\textrm{end}}$ is fixed such that the runtime of each layer is bounded while the initial time $t_{\textrm{init}}$ is parametrized.
(b) Experimental implementation of the parametrized two-qubit gate shown in Fig.~\ref{fig:TwoQubitIdeal}. Because of cross-talk, a direct parametrization of the control parameters of the 2-qubit element has the potential to yield more accurate ground state preparations. Here, the values of the angles $\theta$ and $\phi$ are functions of the coupler parameters $\Phi$ and $t_{\textrm{init}}$. Stray phase shifts $Z_1$ to $Z_4$ are also shown. An alternative implementation of variational $\textrm{iSWAP}(\theta)$ and $\textrm{cphase}(\phi)$ that also uses tunable qubit detunings has been demonstrated in \cite{Foxen2020}.}
\label{fig:TunableCouplerParametrization}
\end{figure}

In the ideal case, we assume to have an explicit mapping between the physical control parameters of the device and the resulting quantum gates in the computational Hilbert space. In the case of perfect controls and characterization of the device, the mapping would be reproducible and invertible as well as perfectly local in the sense that all degrees of freedom can be controlled independently.

In practice, the experimental controls can influence other neighboring gates through residual electromagnetic interactions, namely cross-talk. This means that the mapping between experimental control parameters and the variational angles of the ideal two-qubit gates is not perfectly local. However, for a given assignment of control parameters at a given time, the cross-talk has a reproducible coherent component acting on the computational Hilbert space. Consequently, the variational optimization would benefit from executing the optimization directly on the experimental controls, as it could help mitigate the effect of coherent errors. Fig.~\ref{fig:TunableCouplerParametrization} describes a heuristic scheme to parameterize two-qubit gates on Sycamore at the hardware level. A more detailed implementation has recently been proposed and demonstrated in \cite{Foxen2020}.

In the most general setting, the experimental parametrization can be done over all fluxes and initial times in a given layer. The starting time of the flux pulses $t_{\textrm{init}} \rightarrow \vec{t}_{\textrm{init}}$ and their parametrized amplitudes $\Phi \rightarrow \vec{\Phi}$ become vectors. The angles of the ideal parametrized gates become mappings from the experimental parameters, namely $\vec{\theta}\left(\vec{\Phi},\vec{t}_{\textrm{init}}\right)$ and $\vec{\phi}\left(\vec{\Phi},\vec{t}_{\textrm{init}}\right)$.  The maximum pulse duration $t_{\textrm{end}}$ and amplitude $\Phi_{\textrm{max}}$ are chosen to approximate an $\textrm{iSWAP}$ for a full pulse. The pulses all have the same maximal duration to allow for the composition of layers of variational two-qubit gates that are executed synchronously. The variational single-qubit $Z$ rotations at the end of the two-qubit element can compensate for frequency shifts induced by flux controls. We anticipate that the single-qubit gates can be executed more accurately than the two-qubit gates.

The control pulses we are sending to the device generate gate-like operations. Technological improvements are trending towards the ideal computational gate paradigm of quantum circuits built with independently controlled one- and two-qubit gates \cite{Nielsen2002}. As this paradigm approaches, we gain ansatz interpretability where we can infer a physical interpretation for an assignment of parameters. We also gain ansatz transferability between devices where an assignment of parameters on one device can be transferred to a different device to obtain the same state. 

\begin{figure}[tp]
\centering
\includegraphics[width=8cm]{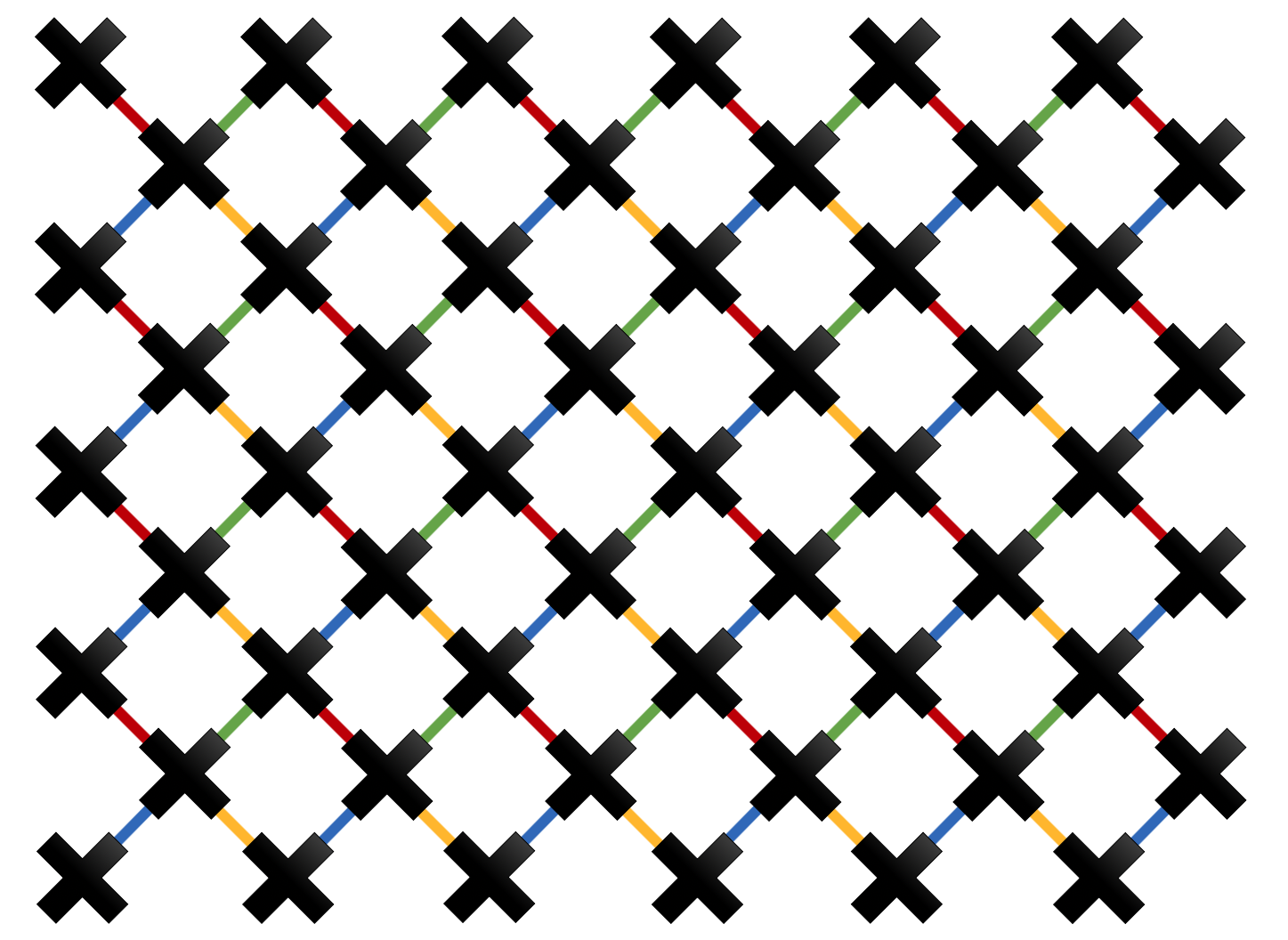}
\caption{Qubit layout and connectivity for the 54-qubit Sycamore chip. Here, edges are classified into 4 sets (A: Red, B: Yellow, C: Green and D: Blue), corresponding to each of the layer patterns composing the Hardware efficient ansatz used in the benchmark.}
\label{fig:SycamoreAnsatz}
\end{figure}

Our ansatz is built by layering staggered patterns of variational 2-qubit gates. Various sequences of patterns can be used. For our numerical examples, the ansatz is constructed by repeating the pattern ABCD as shown in Fig.~\ref{fig:SycamoreAnsatz}. The quantum supremacy demonstration was performed by stacking 20 of these layers \cite{Arute2019}. For VQE applications, the number of layers of the ansatz will typically determine the volume of Hilbert space that can be reached by the variational method. The maximum number of layers is limited by the maximum coherent depth, which is approximately the ratio of the coherence time $T_2$ over the gate time. As described in Section \ref{sec:Implementation}, we can start from a small number of layers and iteratively add new layers until either convergence of the energy is reached or until there is too much noise to improve convergence.



\begin{figure}[tp]
\centering
\subfigure[Interleaved ordering. Horizontal sites. Horizontal chain on sublattices.]{\includegraphics[width=6.5cm]{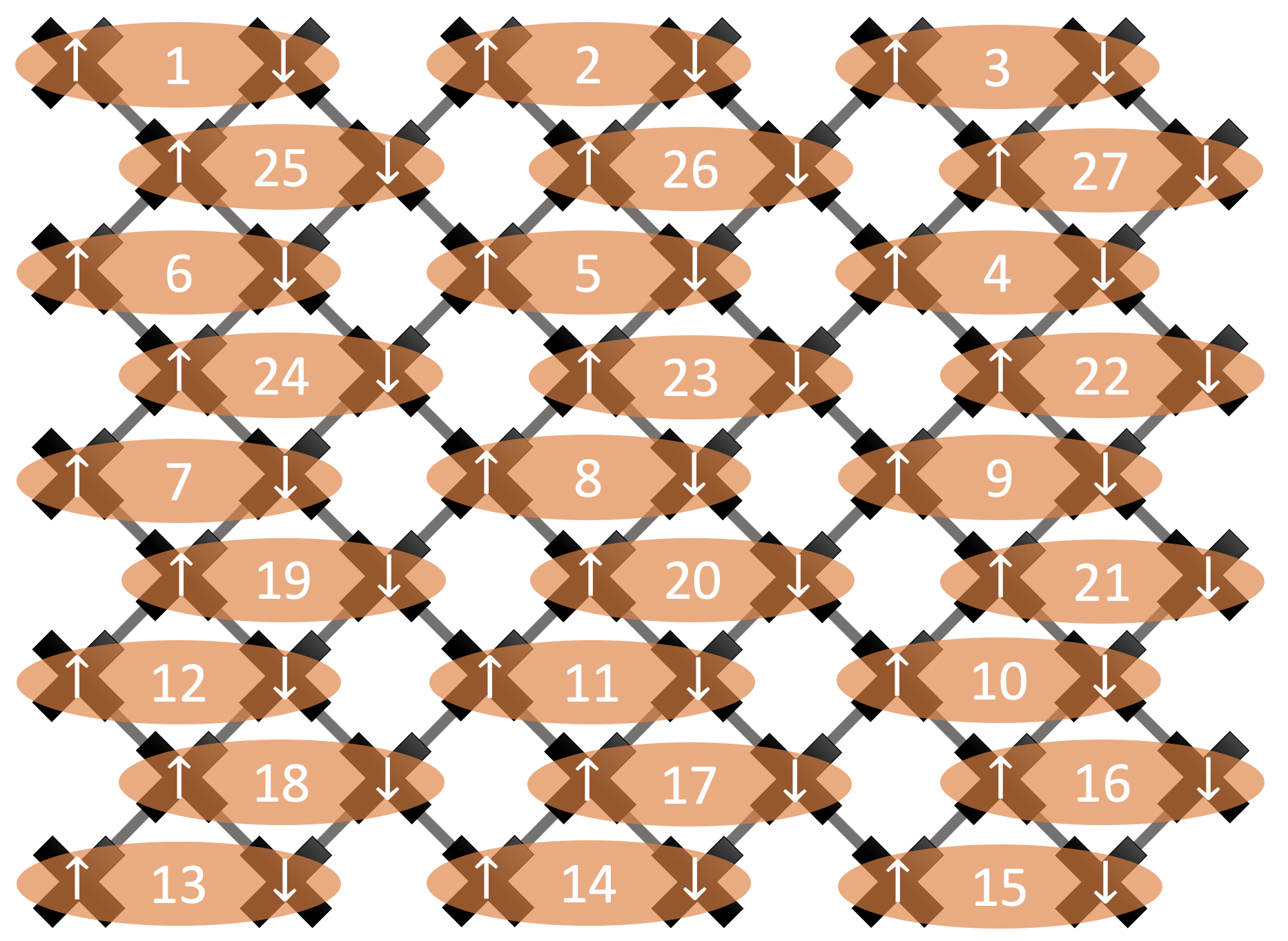}}\\
\subfigure[Vertical ordering. Oblique sites. Vertical chain.]{\includegraphics[width=6.5cm]{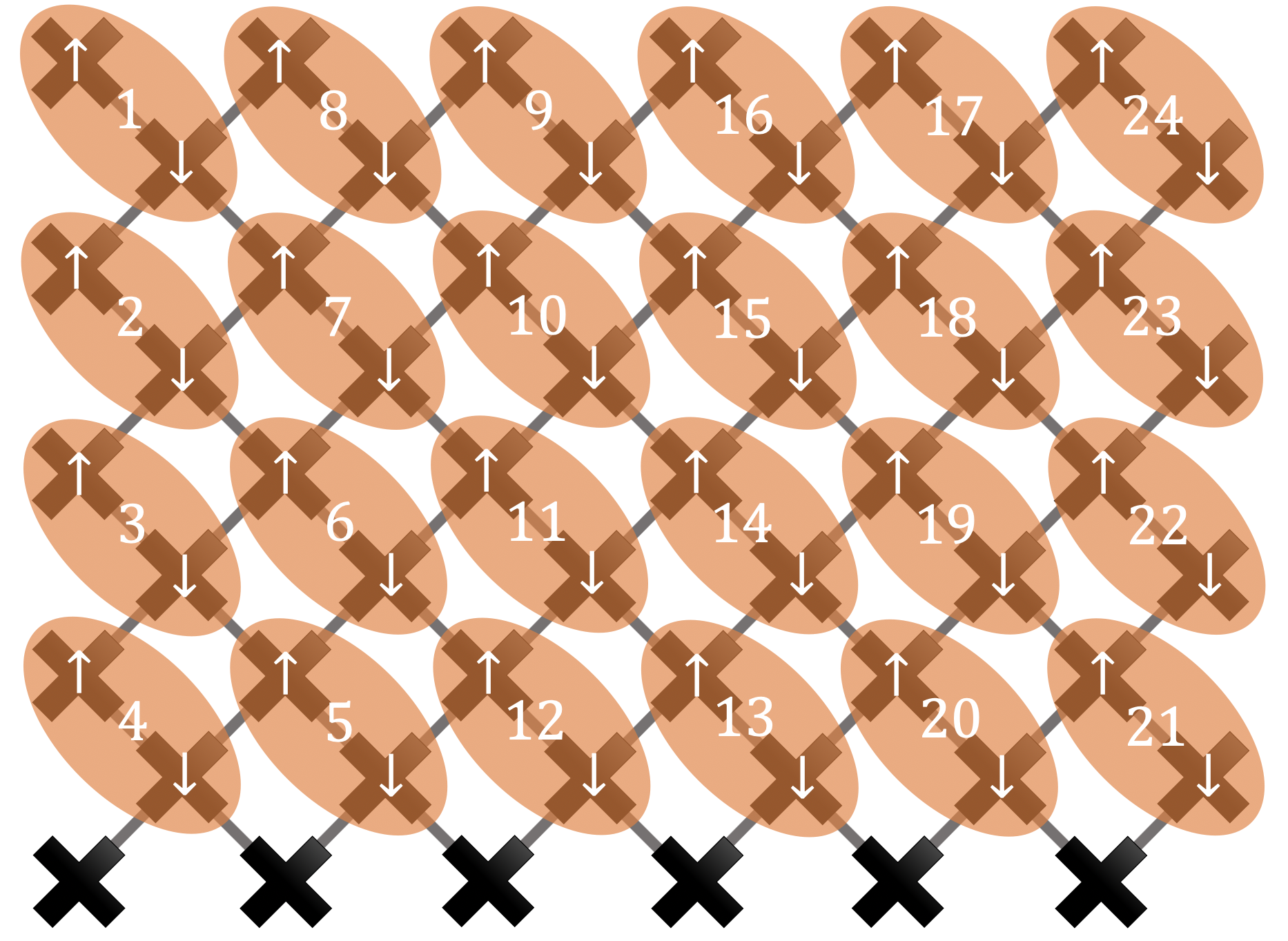}}\\
\subfigure[Horizontal ordering. Oblique sites. Horizontal chain.]{\includegraphics[width=6.5cm]{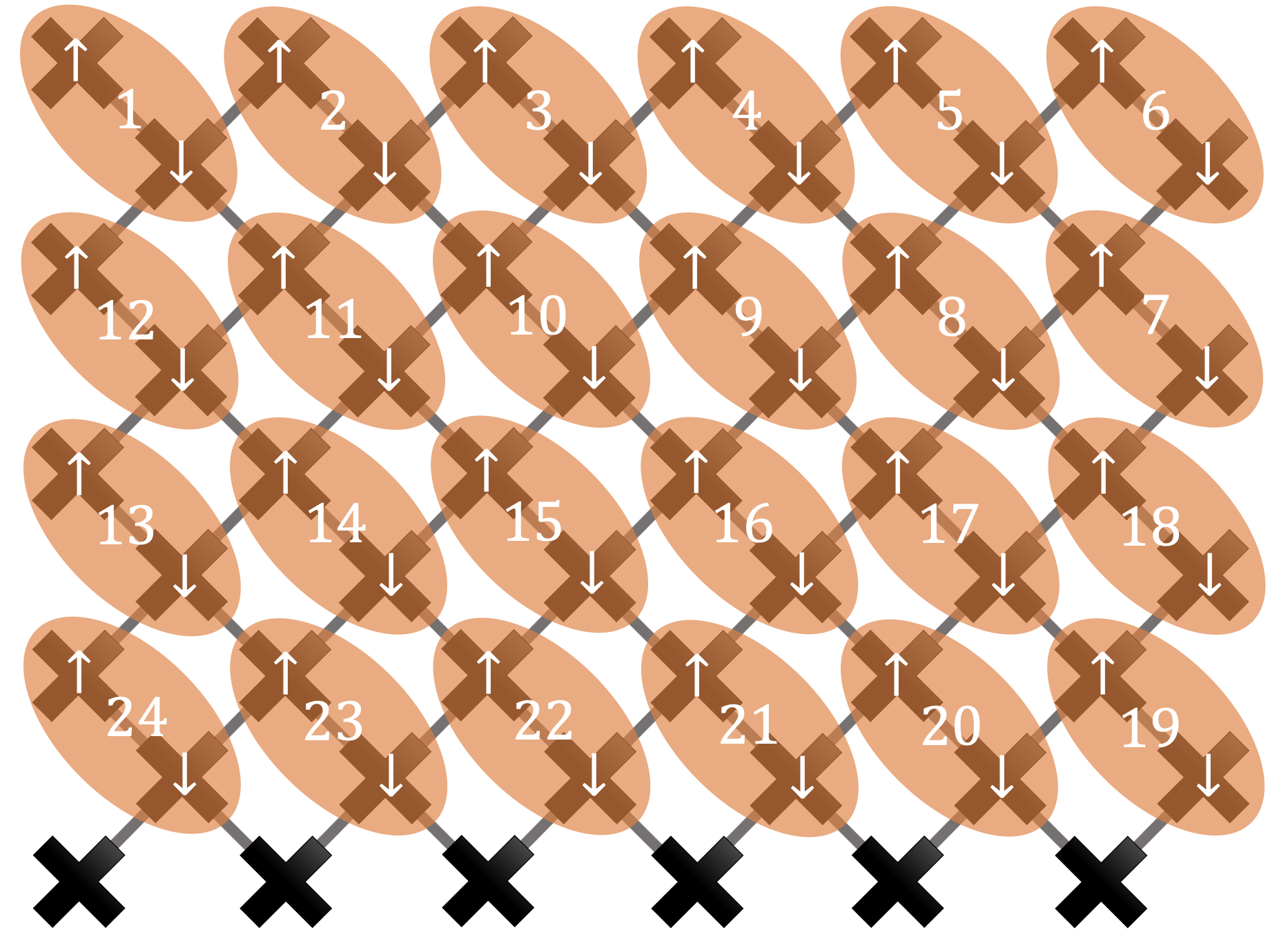}}
\caption{Different circuit ordering (assignments of spin-orbitals to qubit indexes) for the sycamore chip. Each oval corresponds to a site of the Fermi Hubbard model. We assume a canonical Jordan-Wigner ordering for the mapping of spin orbitals in each site.}
\label{fig:SycamoreOrdering}
\end{figure}

\subsubsection*{Fermionic character of the ansatz \label{sub:FermionicCharacter}}

The ansatz generated by the hardware-efficient operations of the Sycamore device is natural for the representation of strongly correlated fermionic states. Along with single-qubit $Z$ rotations, the tunable transverse $XX+YY$ interaction is a generator for matchgates which can be used to construct fermionic gaussian transformations  \cite{Terhal2002,Jozsa2008}. However, non-nearest-neighbor matchgates (such as on a Sycamore lattice) can be used to encode universal quantum computations and are therefore difficult to simulate classically. The variational component $ZZ$ is not part of the set of matchgates either and can also be used to encode a complementary gate set for universal quantum computing. The parametrized $ZZ$ interactions can be interpreted as variational electrostatic terms between fermions.
This means that the bounded-depth Sycamore ansatz can generate a subset of all fermionic Gaussian transformations on a given Jordan-Wigner encoding as well as a non-trivial set of non-Gaussian transformations which are generated by the non-nearest-neighbor variational $XX+YY$ rotations  (on the Jordan-Wigner chain) and the variational $ZZ$ terms between neighboring qubits on the lattice. The permutation of orbitals to obtain a different Jordan-Wigner ordering is itself a fermionic Gaussian transformation which is classically efficient to compute. More details about matchgates and fermionic Gaussian transformation are given in Appendix \ref{sec:Fermionicity}.

\subsubsection*{Hamiltonian and circuit orderings \label{sub:SycamoreOrdering}}

There are two notions of ordering used in the implementation of the proposed ansatz. First, the Jordan-Wigner ordering, and second, the mapping of spin orbitals to physical qubits on the chip.
The first notion was shown in Fig. \ref{fig:JordanWignerEncoding} and it corresponds to the canonical Jordan-Wigner ordering $1\uparrow$,...,$L\uparrow$,$L\downarrow$,...,$1\downarrow$. This ordering determines the Pauli string representation of the Hamiltonian in Eq.~\eqref{eq:onedimensionalFHM_jordanwigner}.
The second notion of ordering is illustrated in Fig.~\ref{fig:SycamoreOrdering} and corresponds to the mapping between simulated spin orbitals and their assigned physical qubit. We will refer to this ordering as \emph{circuit ordering}. We present some possible circuit orderings used in our numerical experiment. In practice, there is a combinatorially large number of possible such orderings and we heuristically chose a few that would approximately preserve the locality of the 1D FHM. As discussed in Section \ref{sec:Numerics}, the EFL can depend on the choice of the ordering, as this impacts the performance of the ansatz. For practical VQE calculations, an optimal ordering can be found with the method described in Appendix \ref{sec:AdvancedTraining}.

\section{Numerical experiments \label{sec:Numerics}}

\begin{figure}[tp!]
\centering
\includegraphics[width=9cm]{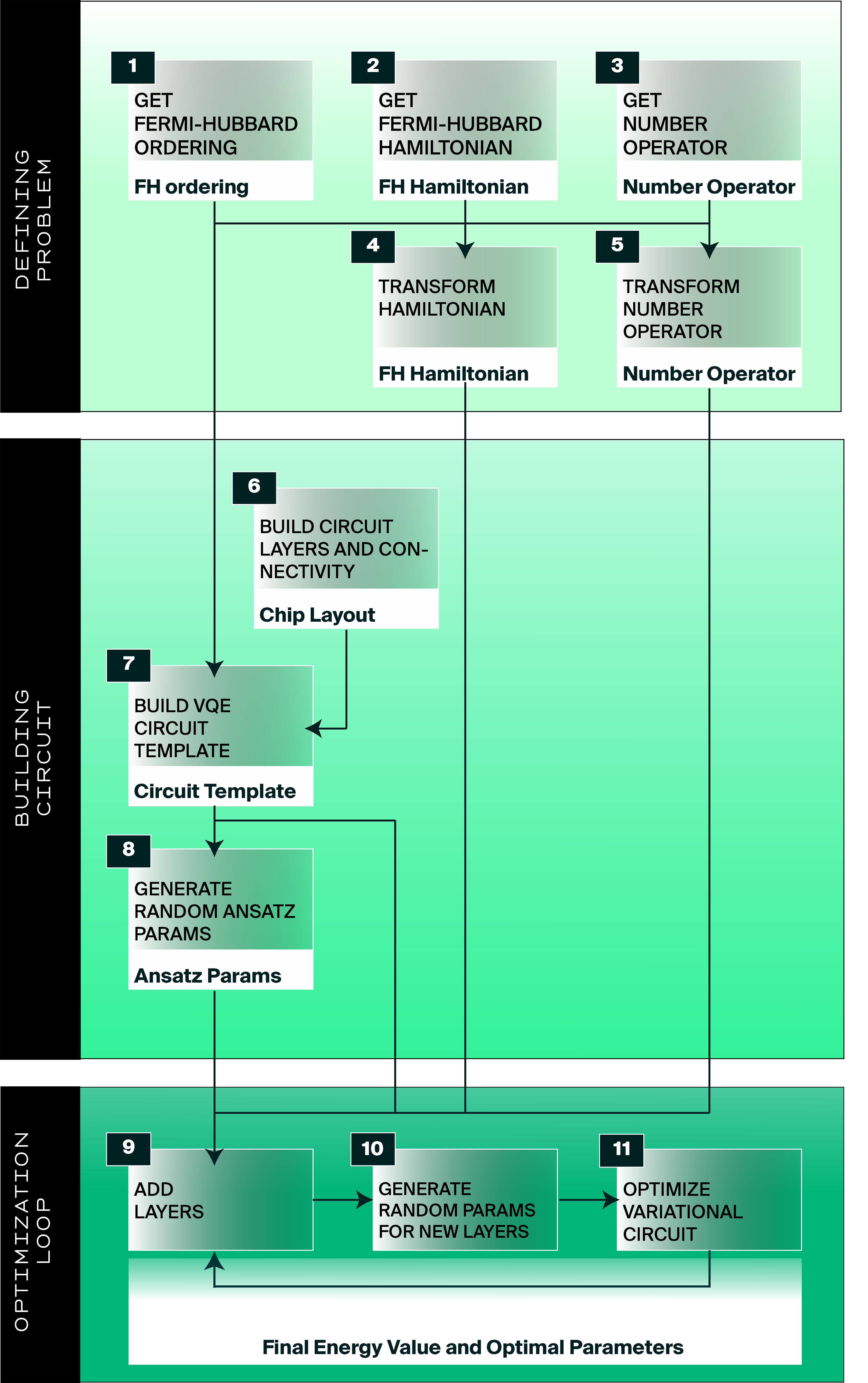}
\caption{Graphical representation of the workflow executing the numerical experiments. Each rectangle represents a step and the associated white box describes the main output of the given step. Each task is executed in a containerized software environment. Arrows represent the flow of data between steps. For clarity, global parameters such as the chain length are not shown. The workflow executes the following sequence of steps: 1) Generation of the circuit ordering; 2) generation of the FH Hamiltonian given the parameters of the system such as chain length, magnetic field, interaction strength and chemical potential ; 3) generation of the number operator used for constraining the number of particles; 4) transformation of the FH Hamiltonian and 5) number operators to qubit representations; 6) generation of the qubit connectivity map corresponding to Sycamore; 7) creation of the VQE circuit template ansatz, which can be then modified by adding layers or using different parameters; 8) generation of random parameters for the first layer of the circuit; 9) addition of layers to the circuit ansatz; 10) generation of random parameters for the new layers; 11) optimization of variational parameters with SLSQP.}
\label{fig:workflow}
\end{figure}

We proceed to test the ideas and methods introduced earlier with numerical simulations. In Section \ref{sub:algorithm_description} we describe the computational workflows used for our numerics and we demonstrate an implementation of the benchmark in Section \ref{sub:OptimalParameters}, along with a discussion of the reproducibility of the proposed layer-by-layer training method.

\subsection{Algorithm description \label{sub:algorithm_description}}
Our numerical simulations were implemented and executed using the Orquestra\textsuperscript{TM} platform built by Zapata Computing \cite{OrquestraWebsite2020}. Orquestra facilitates the execution of experiments that are both scalable and reproducible through workflows. Each workflow defines a sequence of elementary steps involved in the experiments and the inter-dependencies between steps. By dividing an experiment into its constituent steps, independent steps can be easily parallelized. Furthermore, this facilitates the proper allocation of quantum and classical computational resources. A visual depiction of the workflow used for our calculations alongside a detailed description can be found in Fig. \ref{fig:workflow}. We used the Intel Quantum Simulator \cite{Guerreschi2020} to simulate noiseless quantum circuits. Our estimate of expectation values does not include the effect of finite sampling. In all the simulations, we have used the physical parameters $U = 8$, $\mu = 4$ and $t=1$.

\subsection{Demonstration of the benchmark \label{sub:OptimalParameters}}

\begin{figure}[tp!]
\centering
\includegraphics[width=9cm]{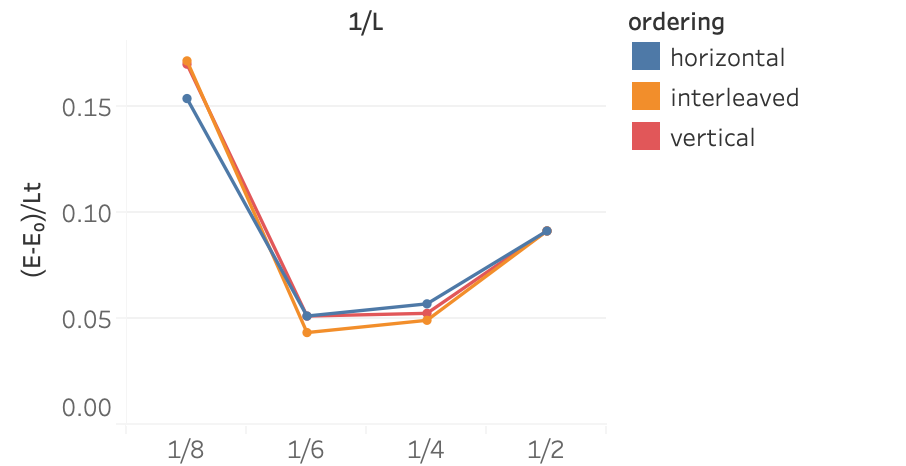}
\caption{Numerical simulations of the Fermionic length benchmark obtained for 1D FHM with lengths 2, 4, 6 and 8, using a Sycamore qubit grid with a bounded depth. The results reproduce the expected behavior described in Fig.~\ref{fig:TheoreticalPerformanceCurve}. Errors in the infinite chain energy density, in the $x$ axis, are estimated from the final energies obtained from the layer-by-layer VQE optimization depicted in Fig.~\ref{fig:convergence_plot}. The maximum circuit depth corresponds to 33 layers of the Sycamore hardware efficient ansatz proposed in Section \ref{sec:Implementation}. For this depth, we find $L^* = 6$. At the circuit depth considered in the simulation, we observe better performance of the interleaved layout .}
\label{fig:performance_curve}
\end{figure}

To illustrate the implementation and execution of our benchmark and how Fig.~\ref{fig:TheoreticalPerformanceCurve} can be obtained in practice, we have performed simulations for different chain lengths. To incorporate the effect of noise in our simulation of the benchmark, Fig.~\ref{fig:TheoreticalPerformanceCurve} was generated with a maximum circuit depth imposed for the simulation. This bounded-depth model operates as a noise model corresponding to having a system with a finite coherence time in a variational circuit. Since the circuit cannot increase as we increase the size of the system, the bounded depth ansatz will perform more poorly as we increase the length of the chain, emulating the effect of decoherence in the error of the infinite chain energy density estimate.

We carried out simulations for chains with length 2, 4, 6 and 8 on qubit grids with a Sycamore connectivity with sizes $2 \times 2$, $2 \times 4$ qubits, $2 \times 6$ and $4 \times 4$, respectively. The benchmark was tested on the circuit ordering shown in Fig.~\ref{fig:SycamoreOrdering}. The VQE optimization is executed in a layer-by-layer fashion, starting with a single layer of the ansatz with initial parameters drawn randomly from the $[0, 2 \pi]$ interval. Then we added 4 layers of the ansatz with initial parameters in the range $[-\frac{\pi}{10}, \frac{\pi}{10}]$ and optimize, limiting the maximum number of optimization steps to 100. This is repeated 8 times, for a total of 32 layers added on top of the initial single layer. We employed the Sequential Least Squares Programming (SLSQP) optimizer \cite{Kraft1988, Kraft1994}, constraining the number of particles to obtain the desired half-filling of the FHM.

\begin{figure*}
\centering
\includegraphics[width=17cm]{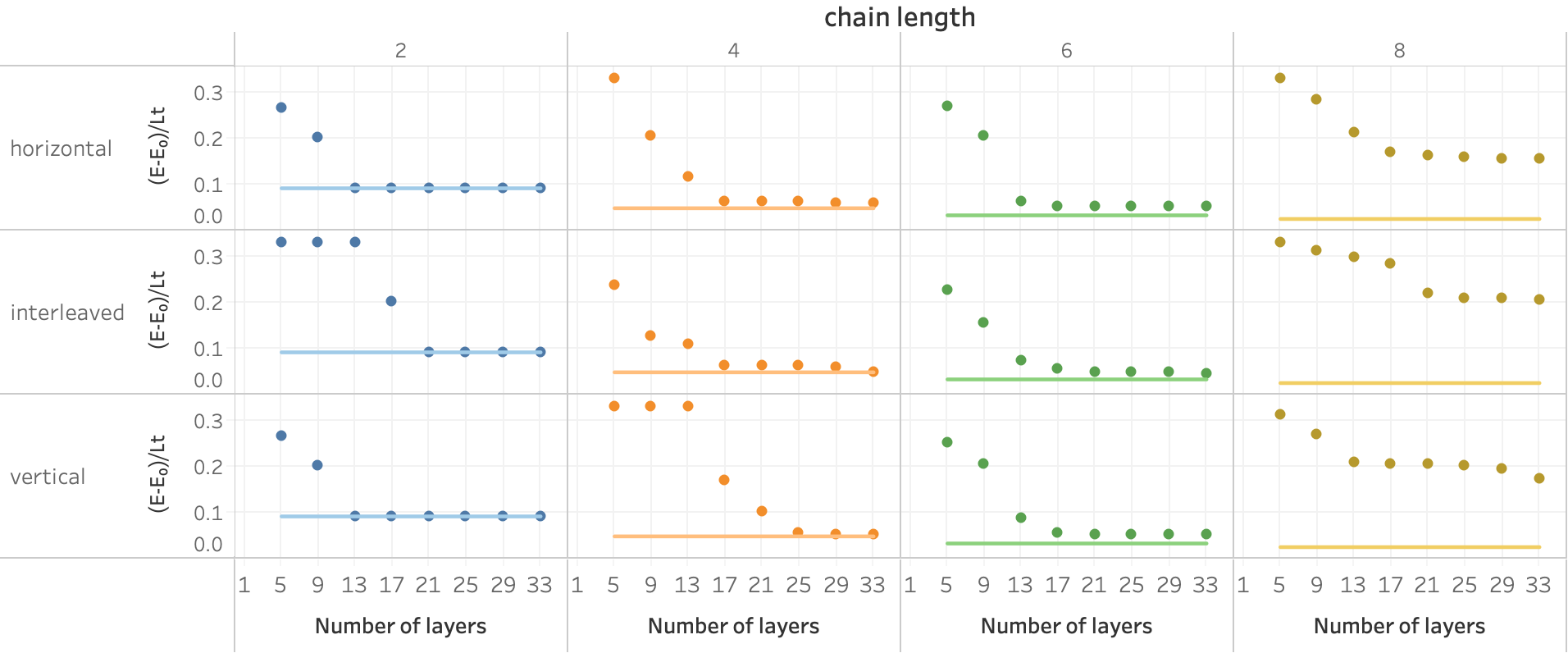}
\caption{Convergence of the error in the infinite chain energy density throughout the layer-by-layer VQE optimization. Each point corresponds to the addition of new layers. The solid lines represent the exact value of the energy. For lengths 2, 4, 6 and 8, the sizes of the qubit grids are $2 \times 2$, $2 \times 4$, $2 \times 6$ and $4 \times 4$, respectively.}
\label{fig:convergence_plot}
\end{figure*}

The benchmark plots obtained for different circuit orderings are presented in Fig.~\ref{fig:performance_curve}. For the maximum circuit depth chosen for the Sycamore ansatz, we achieve convergence close to the ground state energy of the 1D FHM for chains of lengths up to six with all the orderings described in Fig.  \ref{fig:SycamoreOrdering}. We observe that the interleaved layout achieved the best performance for the circuit depth considered in the numerical experiments. Regardless of the ordering, all the calculations show a decreasing trend in the energy density deviation up to chains of length 6. This decreasing trend is interrupted for the $1 \times 8$ chain, for which the maximum circuit depth imposed on the ansatz prevents the VQE calculation to keep improving the estimate of the infinite chain energy density. We therefore conclude that the fermionic length of the Sycamore ansatz with a maximum of 33 layers is $L^* = 6$. Fig.~\ref{fig:convergence_plot} offers details of the energy convergence in the layer-by-layer optimization. For chains of length 1 to 6, the number of layers required for convergence increases with the size of the system and is smaller than the maximum depth allowed. For the $1 \times 8$ chain, the maximum depth does not suffice to attain convergence close to the ground state. The exact ground states were computed using exact diagonalization for reference. 

\subsubsection*{Reproducibility of training results \label{sub:reproducibility}}

\begin{figure}[bp]
\centering
\includegraphics[width=8cm]{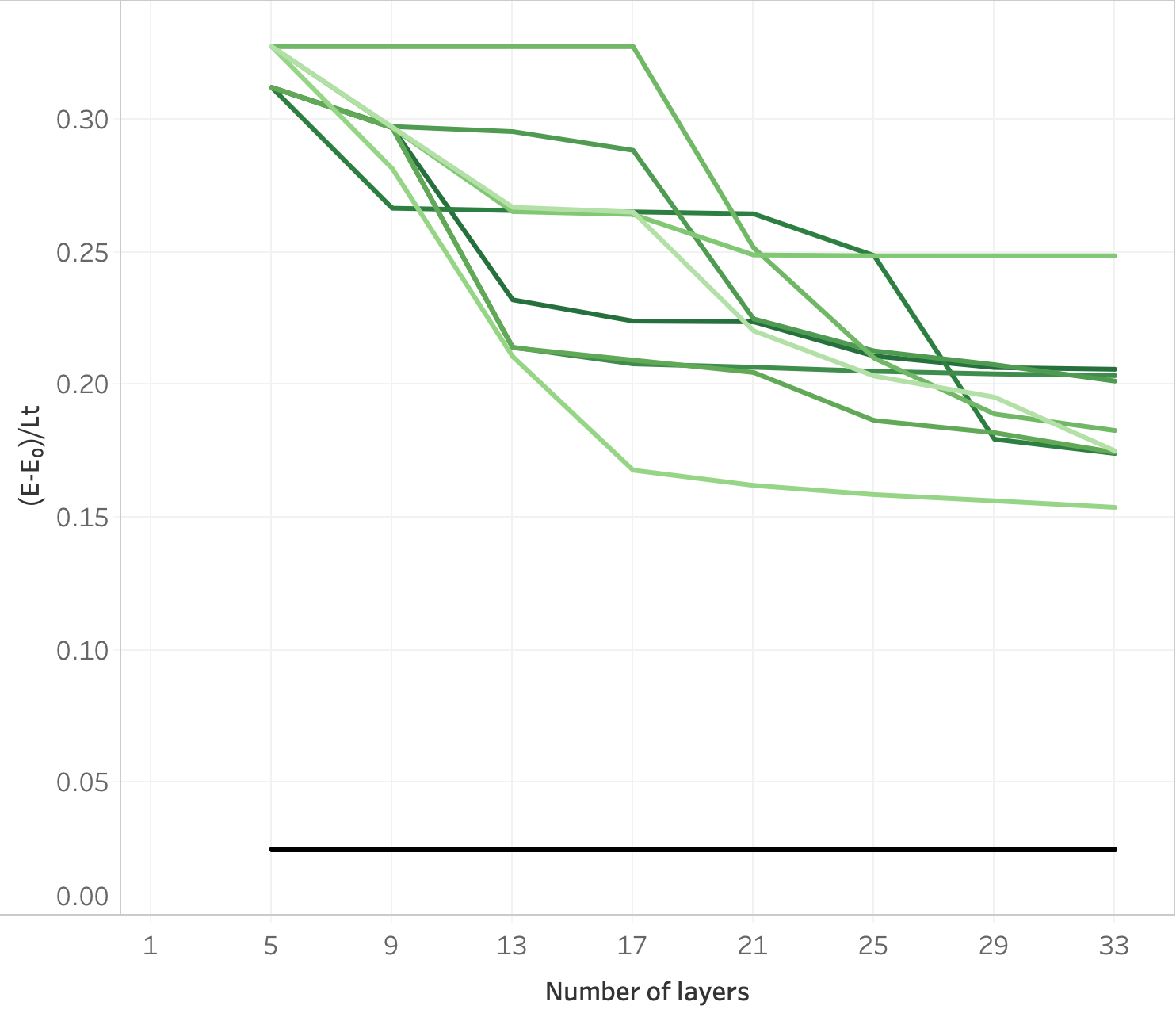}
\caption{Convergence plots for 9 different runs of the fermionic length benchmark for chain of length 8 on a $4 \times 4$ horizontal qubit grid. Each green line represents one run. The black line represents the exact energy for this chain obtained from exact diagonalization. In all cases, the optimization procedure reduces the error in the energy density but do not achieve the optimal value.}
\label{fig:reproducibility}
\end{figure}

To study the sensitivity of the VQE optimization to the random initialization of parameters in our layer-by-layer strategy, we performed 9 VQE simulations with randomly chosen initial parameters of the $1 \times 8$ chain on a $4 \times 4$ grid of qubits with horizontal ordering. As observed in Fig.~\ref{fig:reproducibility}, the best value achieved for the energy density deviation is 0.15, with most of the final values for different calculations concentrating around 0.18. This result highlights how the layer-by-layer strategy consistently decreases the energy throughout the optimization, achieving roughly the same results for all runs. Therefore, the fact that the error expected for the ground state energy is not attained is attributed to the limited circuit depth imposed on the ansatz. In contrast, most of the simulations for the $1 \times 6$ chain and smaller systems converged within an error of 0.02 from the exact value despite slight differences in the convergence paths for different runs.

\section{Discussion \label{sec:Discussion}}

We have proposed a benchmark to characterize the ability of near-term quantum processors to simulate a fermionic system with the variational quantum eigensolver. Our application benchmark utilizes the 1D FHM as the test Hamiltonian, providing an effective fermionic length that quantifies the size of fermionic systems that can be studied in a device for a particular choice of variational ansatz. Our benchmark is scalable in the sense that it requires a modest number of measurements and the knowledge of the exact energy of the 1D FHM in the thermodynamic limit, which can be obtained analytically. We provide a concrete implementation proposal of the benchmark for the Sycamore processor \cite{Arute2019} and demonstrate its viability through numerical experiments. Our results show that the hypothesized behavior of the benchmark is recovered when considering variational circuits with bounded depth as a way to emulate the effect of noise. 

An important aspect of our benchmark is its runtime requirement, which depends on the number of measurements utilized in the VQE energy estimation process. The sampling rate is independent of the number of qubits and is inversely proportional to the depth of the circuit on the experimental device. On the Sycamore device, samples are obtained at a rate of 5 kHz. Assuming that measurements are distributed according to an optimal operator averaging strategy \cite{marginal_constraints}, and taking advantage of grouping of co-measurable terms, we have estimated that reaching an accuracy of $10^{-2}$ for the energy density of the $1 \times 2$ chain would require about 9 seconds of sampling, while an accuracy of $10^{-3}$ would take around 15 minutes. Here, we also assume that covariances among the terms in the same group average to zero, which in our experience results in a slight overestimation of the number of necessary measurements. As pointed out in Section \ref{sub:VQETask}, the number of measurements, and therefore estimation times, are expected to \emph{decrease} for maintaining the same error in estimating energy {per site} as the chain gets longer. This analysis indicates that estimating the required quantities for the benchmark is practical on existing quantum devices.

A potential obstacle to the scalability of our scheme lies in the number of function evaluations required for training the ansatz, which is not in the scope of this paper but for which we recognize ample opportunities for further improvements. A significant improvement of the training method could involve a more advanced initialization method of the parameters using matrix product states \cite{Ran2019} and tensor networks \cite{Biamonte2017} for arbitrary universal hardware-efficient ansatzes. A simpler training heuristic could also involve using adiabatic-assisted VQE \cite{Garcia2018} to construct increasingly longer chains of the FHM. Similarly, characterizing the impact of different device errors in the VQE optimization, and simulations of the benchmark under more realistic noise conditions are left to future work.

Another possible avenue for improvement relates to the fermionic nature of the device ansatz as described in Section \ref{sub:FermionicCharacter}. As shown in Ref.~\cite{Dallaire2019}, a circuit consisting of nearest-neighbor iSWAP gates corresponds to a basis rotation realized by time evolution under a free-fermion Hamiltonian \cite{Terhal2002}. Therefore one could improve the training by first using classical mean-field calculations to prepare the optimal Gaussian state using a basis rotation, and proceed to add parameterized ZZ gates and non-nearest-neighbor iSWAP gates for refining the ansatz into the space of non-Gaussian states \cite{Dallaire2019}.
Once the benchmark has been executed on a device and found useful for a given application of VQE, a more refined training method would be the one described in Appendix \ref{sec:AdvancedTraining}. This method has the advantage that it cannot yield a state with higher energy than the mean-field state, which can be computed efficiently on a classical computer but it generally requires more measurements to converge than the 1D FHM benchmark.

Future research will explore the error resilience \cite{sharma2019noise} of the proposed ansatz. In particular, we hypothesize that just as some coherent errors can be mitigated by a parameterized quantum circuit, the incorporation of auxiliary qubits could mitigate some incoherent errors as they allow for the implementation of parameterized quantum channels. In this case, the auxiliary qubits serve effectively as a bath such as the entropy generated by incoherent processes can be dissipated through them, opening a potential avenue to improve the performance of VQE. The study of this hypothesis as well as the interplay of the proposed benchmark with previously proposed error mitigation techniques \cite{temme2017,endo2018practical,kandala2019error,mcardle2018error} are left for future work. 

Another research direction is the use of the VQE ansatz and optimization strategy proposed in this paper in the study of more complex fermionic simulation problems. For instance, as the computation of the Green$'$s function of the one-dimensional FHM is difficult in practice with the Bethe ansatz, we propose using subspace-search VQE to measure the single-particle Green$'$s function of the Hubbard model as is done in \cite{Endo2019}. Also, once a one-dimensional FHM chain can be converged on a quantum processor, we propose using adiabatic-assisted VQE to prepare the ground state of more complicated systems like the two-dimensional FHM. Furthermore, it would be possible to try optimizing a $3 \times 3 \times 3$ instance of the FHM by taking advantage of the full size of the Sycamore chip. This would be a step on the path to concrete applications of quantum computing technologies.

Finally, it should be possible to design hardware efficient fermionic ansatzes for other quantum computing architectures such as ion traps. In principle, this ansatz could be realized using native gates, such as the M{\o}lmer-S{\o}rensen gate, and a trap architecture that allows for the execution of a number of simultaneous two-qubit gates that scales with the number of qubits.

\section*{Acknowledgement \label{Acknowledgement}}

PLDD proposed the theoretical aspects of the benchmark. JG, JR, NTB and MS composed the Orquestra workflows to generate the simulation data. MS and NTB executed the workflows and analyzed the results. PLDD, JR and YC wrote the paper. The authors thank Peter Johnson for inspiration of the advanced training method as well as Max Radin, Borja Peropadre, Morten Kjaergaard and Gabriel Samach for helpful discussions and insights, and Maria Genckel for the much improved presentation of the workflow diagram. We also thank Jonny Olson for proof-reading and fine-tuning the final manuscript.

\bibliography{onedfhm}

\appendix

\section{Fermionic character of the ansatz \label{sec:Fermionicity}}

Here we present a short review of the formalism of fermionic gaussian transformations and their correspondence to networks of nearest-neighbor matchgates. For a more detailed treatment we refer to \cite{Terhal2002,Jozsa2008}.
It is useful to use the real quadrature of the fermionic field which are the Majorana operators
\begin{equation}
    \begin{array}{rcl}
     \gamma_k^A&=&a_k^{\dagger}+a_k\\
     &&\\
     \gamma_k^B&=&-i(a_k^{\dagger}-a_k).
    \end{array}
    \label{eq:FermionicQuadratures}
\end{equation}
In the diagonal basis of a quadratic fermionic Hamiltonian, a Slater determinant over $M$ spin orbitals can be constructed from the vacuum as a product of single particles
\begin{equation}
    \left|\Phi_0\right\rangle = \prod_{k=1}^{L} (\gamma_k^A)^{\frac{1+\lambda_k}{2}}\left|\textrm{vac}\right\rangle
    \label{eq:InitialProductState}
\end{equation}
where the $\lambda_k$ are the Williamson eigenvalues. In general, the density matrix of a fermionic gaussian state can always be represented in the form
\begin{equation}
    \rho = \frac{1}{2^L}\prod_{k=1}^L\left(1+i\lambda_k\gamma_k^A\gamma_k^B\right).
    \label{eq:GeneralFermionicGaussianState}
\end{equation}
A more compact notation is given by a $2M \times 2M$ covariance matrix whose elements can be computed as
\begin{equation}
    \Gamma_{kl} = \frac{i}{2}\textrm{tr}\left(\rho[\gamma_k,\gamma_l]\right).
    \label{eq:CovarianceMatrix}
\end{equation}
This can be written in the diagonal form as
\begin{equation}
    \Gamma = R \bigoplus_k^L \begin{pmatrix} 0&-\lambda_k\\\lambda_k&0\end{pmatrix} R^{\intercal}
    \label{eq:DiagonalFormCovarianceMatrix}
\end{equation}
where $R$ is a $\textrm{SO}(2L)$ rotation. Those rotations are Bogoliubov transformations in general and orbital rotations in the particular case where the number of particles is conserved.

We can now establish the connection to nearest-neighbor matchgates introduced by Valiant \cite{Valiant2002,Valiant2008}. A general matchgate $G$ between two qubits is defined by
\begin{equation}
    G(A,B) = \begin{pmatrix} p&0&0&q\\ 0&w&x&0\\ 0&y&z&0\\ r&0&0&s\end{pmatrix},
    \label{eq:Matchgate}
\end{equation}
where $A = \begin{pmatrix} p&q\\ r&s\end{pmatrix}$ and $B = \begin{pmatrix} w&x\\ y&z\end{pmatrix}$ are $\textrm{SU(2)}$ rotations with the same determinant $\textrm{det}(A)=\textrm{det}(B)$. Nearest-neighbor matchgates acting on two qubits have six generators
\begin{equation}
    \begin{array}{rcl}
    X_j X_{j+1}&=& -i \gamma_{j}^B \gamma_{j+1}^A\\
    &&\\
    X_j Y_{j+1}&=& -i \gamma_{j}^B \gamma_{j+1}^B\\
    &&\\
    Y_j X_{j+1}&=& i \gamma_{j}^A \gamma_{j+1}^A\\
    &&\\
    Y_j Y_{j+1}&=& i \gamma_{j}^A \gamma_{j+1}^B\\
    &&\\
    Z_j&=& -i \gamma_{j}^A \gamma_{j}^B\\
    &&\\
    Z_{j+1}&=& -i \gamma_{j+1}^A \gamma_{j+1}^B
    \end{array}
    \label{eq:MatchgateGenerators}
\end{equation}
which explicitly correspond to quadratic forms of Majorana operators after a Jordan-Wigner transformation. In general, a full network of matchgates can generate any $\textrm{SO}(2L)$ rotation. A general rotation is parametrized by $2L^2-L$ angles and can be implemented in linear circuit depth on a linear array of qubits \cite{Dallaire2019}.

If $V$ is a circuit of nearest-neighbor matchgates, the holographic relation that translates the transformation from Hilbert space to a rotation in the space of fermionic operators is given by
\begin{equation}
    V \gamma_j V^{\dagger} = \sum_{k=1}^{2L} R_{kj} \gamma_k.
    \label{eq:HolographicRelation}
\end{equation}

Finally, the connection to fermionic gaussian transformation is established through quadratic fermionic Hamiltonian of the form
\begin{equation}
    H = i \sum_{j \neq k = 0}^{2L-1} h_{jk} \gamma_j \gamma_k,
    \label{eq:QudraticHamiltonian}
\end{equation}
where $h$ is a real and antisymmetric matrix. A general $\textrm{SO}(2L)$ Bogoliubov transformation in the space of fermionic ladder operators has the form $R=e^{4h}$.

\section{Advanced training method \label{sec:AdvancedTraining}}

The ansatz defined in Section \ref{sec:Implementation} can be used for the general VQE procedure. In that case, we are interested to find a fermionic ground state which improves upon the mean field state which is classically efficient to compute. To maximize the use of the non-Gaussian resources contained in the quantum ansatz, it is best to offload all Gaussian transformations with a classical computer.

The full cost function for an ansatz that do not necessarily conserve the number of particles is given by
\begin{equation}
    \Omega(\vec{\theta},\vec{\phi})=\textrm{tr}\left(\left(H(\vec{\phi})-\mu N(\vec{\phi})\right)\rho(\vec{\theta}) \right)
    \label{eq:CostFunction}
\end{equation}
where $\rho(\vec{\theta})=U(\vec{\theta})\left|0\right\rangle\left\langle 0\right| U(\vec{\theta})^{\dagger}$ is the state parametrized on the quantum computer and $H(\vec{\phi}) = V(\vec{\phi})H V(\vec{\phi})^{\dagger} = R_{\vec{\phi}}(H)$ is the Hamiltonian represented in a basis which can be optimized variationally on a classical computer.

The general training method is a modified Newton procedure where at each step $k$ we compute the cost function
At each step, we must compute the cost function $\Omega\left(\vec{\theta}^{(k)},\vec{\phi}^{(k)}\right)$. The parameters are updated with the rule 
\begin{equation}
    \begin{array}{rcl}
    \vec{\phi}^{(k+1)}&=&\vec{\phi}^{(k)}+\Delta\vec{\phi}^{(k)}\\
    &&\\
    \vec{\theta}^{(k+1)}&=&\vec{\theta}^{(k)}+\Delta\vec{\theta}^{(k)},
    \end{array}
    \label{eq:UpdateRule}
\end{equation}
where $\Delta\vec{\theta}^{(k)}$ and $\Delta\vec{\phi}^{(k)}$ are computed from
\begin{equation}
    \begin{pmatrix} \Delta\vec{\phi}^{(k)} \\ \Delta\vec{\theta}^{(k)}\end{pmatrix} = -\left(B^{(k)}\right)^{-1}\nabla \Omega(\vec{\theta}^{(k)},\vec{\phi}^{(k)}).
    \label{eq:NewtonStep}
\end{equation}We define the global gradient operator as $\nabla = \begin{pmatrix} \nabla_{\vec{\phi}}\\ \nabla_{\vec{\theta}}\end{pmatrix}$ and the Hessian as $B = \nabla \nabla^{\intercal}\Omega(\vec{\theta},\vec{\phi})$. The gradients can be measured as
\begin{equation}
    \begin{array}{rcl}
    \nabla_{\phi_k}\Omega(\vec{\theta},\vec{\phi}) &=& i \textrm{tr}\left(R_{k-1:0} \circ d_{h_k}\circ R_{M:k}(\Omega)\rho(\vec{\theta})\right)\\
    &&\\
    \nabla_{\theta_k}\Omega(\vec{\theta},\vec{\phi})&=& i\textrm{tr}\left(\Omega(\vec{\phi})\nabla_{\theta_k}\rho(\vec{\theta})\right)
    \end{array}
    \label{eq:CostFunctionGradients}
\end{equation}
where $d_{h_k}(X)=[X,h_k]$. The second order derivative have the form
\begin{widetext}
\begin{equation}
    \begin{array}{rcl}
    \nabla_{\phi_j}\nabla_{\phi_k}^{\intercal}\Omega(\vec{\theta},\vec{\phi}) &=& -\textrm{tr}\left(R_{j-1:0}\circ d_{h_j}\circ R_{k-1:j}\circ d_{h_k}\circ R_{M:k}(\Omega)\rho(\vec{\theta})\right)\\
    &&\\
    \nabla_{\theta_j}\nabla_{\theta_k}^{\intercal}\Omega(\vec{\theta},\vec{\phi}) &=& -\textrm{tr}\left(\Omega(\vec{\phi}) \nabla_{\theta_j}\nabla_{\theta_k}^{\intercal}\rho(\vec{\theta})\right) \\
    &&\\
    \nabla_{\phi_j}\nabla_{\theta_k}^{\intercal}\Omega(\vec{\theta},\vec{\phi}) &=& -\textrm{tr}\left(R_{j-1:0}\circ d_{h_j}\circ R_{M:j}(\Omega)\nabla_{\theta_k}^{\intercal}\rho(\vec{\theta})\right).
    \end{array}
    \label{eq:CostFunctionHessian}
\end{equation}
\end{widetext}
The first and second derivatives with respect to $\vec{\theta}$ can be measured with the parameter shift rule \cite{Schuld2019Gradient}.

\end{document}